\documentclass[preprint,12pt]{elsarticle} 

\usepackage{amsmath}
\usepackage{listings}
\usepackage{color}
\usepackage{algpseudocode}
\usepackage{algorithm}
\usepackage{empheq}
\usepackage[most]{tcolorbox}

 \usepackage{graphics}

 \usepackage{graphicx}
\usepackage{csquotes}
\usepackage{amssymb}

\newcounter{bla}

\journal{Journal of Computational Science}

\begin{document}
\lstset{language=[90]Fortran,
  basicstyle=\ttfamily,
  keywordstyle=\color{blue},
  commentstyle=\color{magenta},
  morecomment=[l]{!\ }
}

\begin{frontmatter}

\title{Thread-safe lattice Boltzmann for high-performance computing on GPUs}

\author[a]{Andrea Montessori\corref{author}}
\author[b]{Marco Lauricella}
\author[b]{Adriano Tiribocchi}
\author[c]{Mihir Durve}
\author[a]{Michele La Rocca}
\author[d]{Giorgio Amati}
\author[e]{Fabio Bonaccorso}
\author[c,f]{Sauro Succi}

\cortext[author] {Corresponding author.\\\textit{E-mail address:} andrea.montessori@uniroma3.it}
\address[a]{Dipartimento di Ingegneria Civile, Informatica e delle Tecnologie Aeronautiche, Università degli Studi Roma TRE, via Vito Volterra 62, Rome, 00146, Italy}
\address[b]{Istituto per le Applicazioni del Calcolo CNR, via dei Taurini 19, 00185 Rome, Italy}
\address[c]{Center for Life Nano Science@La Sapienza, Istituto Italiano di Tecnologia, 00161 Roma, Italy}
\address[d]{SCAI, SuperComputing Applications and Innovation Department, CINECA, Via dei Tizii, 6, Rome 00185, Italy}
\address[e]{Department of Physics and INFN, University of Rome Tor Vergata, Via della Ricerca Scientifica 1,
00133 Rome, Italy}
\address[f]{Department of Physics, Harvard University, Cambridge, MA, 02138, USA}

\begin{abstract}

We present thread-safe, highly-optimized lattice Boltzmann implementations, specifically aimed at exploiting the high memory bandwidth of GPU-based architectures. At variance with standard approaches to LB coding, the proposed strategy, based on the reconstruction of the post-collision distribution via Hermite projection, enforces data locality and avoids the onset of memory dependencies, which may arise during the propagation step, with no need to resort to more complex streaming strategies. 
The thread-safe lattice Boltzmann achieves peak performances, both in two and three dimensions and it allows to sensibly reduce the allocated memory ( tens of GigaBytes for order billions lattice nodes simulations)  by retaining the algorithmic simplicity of standard LB computing.
Our findings open attractive prospects for high-performance simulations of complex flows on GPU-based architectures.

\end{abstract}


\end{frontmatter}

\section{Introduction}

In the last two decades, the lattice Boltzmann (LB) method has gained a prominent role in the CFD community due to its conceptual and practical simplicity and to the built-in, strongly appealing computational features which can be compactly summarized in the well-known mantra \textit{non-locality (streaming) is linear and non-linearity (collision) is local} \cite{succi}. These two features (among others) grant the success of LB over more established, standard approaches based on the direct discretization of the Navier-Stokes equations in which the convective term $\mathbf{u} \cdot \nabla \mathbf{u}$ (being $\mathbf{u}$ the fluid velocity), non-linear and non-local at a time, becomes particularly demanding to integrate when dealing with complex flows, such as fluid turbulence or flows in complex geometries.
On the other hand, the LB is known to possess a relatively low operational (or arithmetic) intensity, in the range $1 \to 5$ FLOPs/byte (floating point operations per second per byte of memory accessed), depending on the collision strategy and/or the floating point representation employed \cite{exasc2}. Such a low operational intensity prevents to exploit the ideal peak performance which could be potentially delivered by processors and graphic accelerators.
Thus, from a computational standpoint, it is clear that the fundamental interest in LB computing lies in the development  of effective strategies to mitigate the impact of data access and to reduce memory occupancy, a practice that is particularly recommended in GPU-based computing.  
Leaving aside the use of lighter-weight, customized floating point data representations, two fundamental attempts which have been proposed to realize optimal LB implementations  on GPU are represented by the streaming pattern of Bailey \cite{bailey2009accelerating} and the in-place streaming of Geier \cite{geier2017esoteric}, the so-called \textit{Esoteric twist}. Both allow to perform LB simulations with just one set of distribution functions (per fluid component) allocated in RAM, thus halving the memory occupancy of LB applications. On the other hand, such memory saving does not come for free. First of all, both the Bailey streaming and the \textit{Esoteric twist} break the intrinsic simplicity of the lattice Boltzmann, since the  distributions perform different operations depending on the parity of the LB step \cite{bailey2009accelerating}. Moreover, such in-place streamings are \enquote{language-dependent} since their implementations require the resort to object-oriented programming.  More precisely, both approaches relies upon the swapping of data pointers between even and odd time steps so to preserve data locality \cite{lehmann2022esoteric}, a fundamental requirement on shared-memory architectures.

The aim of this work is to present a thread-safe implementation of the  LB algorithm, for single and multicomponent flows, aiming at reconstructing the set of post-collision/post-streaming LB distributions on a lattice node by explicitly computing non-equilibrium contributions via Hermite projection \cite{zhang2006efficient}.

Such a thread-safe LB enforces data locality, thus permitting to perform fused streaming and collision implementation by allocating only a set of distributions per fluid species, it avoids the onset of race conditions during the propagation step and it delivers peak performances as shown by the roof-line model.  More importantly still, the beneficial features above can be obtained without the need to opt for more complicated streaming strategies, as the ones cited previously. 
In conclusion, such a thread-safe implementation is simple and, perspectively, it could be easily incorporated into existing HPC softwares with ideally no extra cost in terms of coding complexity.

\section{The lattice Boltzmann equation for complex flows}

In this section, we briefly summarize the main features of the LB equation \cite{succi2018,kruger2017lattice,montessori2018lattice}. 
The LB with single relaxation time approximation reads as follows:

\begin{equation} \label{lbe}
    f_a(\mathbf{x}+\mathbf{c}_a\Delta \tau,\tau + \Delta \tau)=f_a(\mathbf{x},t) + \omega(f^{eq}_a(\rho,\rho\mathbf{u})-f_a(\mathbf{x},t))
\end{equation}

where, $f_a(\mathbf{x},t)$ is the set of distribution functions, $a=1,..,q$ spanning the $q$ lattice vectors $\mathbf{c_a}$ ($q$ depending on the dimensionality and isotropy of the lattice in use \cite{succi2018, montessori2015lattice}), $f^{eq}(\rho,\rho\mathbf{u})$ is a discrete set of thermodynamic equilibria obtained through a second-order Mach expansion of the Maxwell-Boltzmann distribution of the molecular velocity:

\begin{equation} \label{equil}
    f_a^{eq}=t_a\left(\rho + \rho_0\left(\frac{\mathbf{c}_{a} \cdot \mathbf{u}}{c_s^2} + \frac{(\mathbf{c}_a\mathbf{c}_a - c_s^2\mathbb{I}): \mathbf{u}\mathbf{u}}{2c_s^4}   \right )\right).
\end{equation}

The set of equilibria has been chosen as in \cite{he1997lattice} to enforce the incompressibility of the fluid.  In eq.\ref{equil}, $\rho_0$ is the bulk density of the fluid, $\rho=\rho_0 + \delta\rho$ is the total density of the fluid carrying information of the density perturbations $\delta \rho$, $\mathbb{I}$ is the identity second order tensor, the symbol \enquote{$:$} indicates a dyadic product between tensors, $t_a$ is a lattice-dependent set of weights \cite{kruger2017lattice}.

Finally, $\mathbf{x}$ and $\tau$ are, respectively  a three-dimensional vector denoting the physical space and the time.
As evidenced in eq.\ref{lbe}, an LB algorithm consists of two main steps, namely a local (in time and space) collision which can be performed via a single time relaxation, towards a set of local Maxwell-Boltzmann equilibria occurring at a rate proportional to $\omega$, and a free-flight of the set of distributions along the lattice directions. The first operation is local and contains non-linearity in the macroscopic velocity while the second one is linear, non-local and exact at machine precision (no loss of information).
The reconstruction of statistical moments up to a given order is tantamount to reconstruct the main hydrodynamic fields of interest namely, density, linear momentum and momentum flux tensor:

\begin{equation}
    \rho=\sum_a f_a, \;\rho\mathbf{u}=\sum_a f_a \mathbf{c}_a, \; \Pi_{\alpha\beta}=\sum_a f_a c_{a\alpha}c_{a\beta}.
\end{equation}

Through a multi-scale Chapman-Enskog expansion \cite{chapman1990mathematical} in the smallness parameter $\epsilon$ (i.e. Knudsen number), it is possible to show that the lattice Boltzmann equation recovers, macroscopically, a set of  partial differential equations governing the conservation of mass and momentum of a volume of fluid featuring a kinematic viscosity $\nu=c_s^2(1/\omega-0.5)$ (being $c_s^2=1/3$ the speed of sound of the model) i.e.:

\begin{equation}
    \nabla \cdot \mathbf{u}=0,
\end{equation}

\begin{equation}
    \frac{\partial \mathbf{u}}{\partial t} +\mathbf{u}\cdot \nabla \mathbf{u} = -\frac{1}{\rho}\nabla p +\nu \Delta \mathbf{u}.
\end{equation}

The evolution equation eq.\ref{lbe} can be further augmented with additional forcing terms to include the effect of external body forces, such as gravity or electrostatic forces, and with dispersion-like terms mimicking the build-up of a positive surface tension in a system of immiscible fluids (see \cite{kruger2017lattice, succi2018, montessori2018lattice} for further details).

In the next subsection, we describe a possible approach to include surface tension effects within an LB framework.

\subsection{Color-gradient lattice Boltzmann method for multicomponent flows}

In the LB context the effect of a positive surface tension, arising among interacting immiscible fluids, can be modeled by introducing suitable pseudo-potentials, coarse-graining the microscopic interactions among molecules of different fluids \cite{shan1993lattice, swift1996lattice, montessori2017entropic,benzi2006mesoscopic}, or by directly tracking the dynamic evolution of the fluid interface \cite{anderl2014free, thommes2009lattice}. 
Among such approaches, the color-gradient (CG) method has proven to be a fruitful choice given its ease of implementation, stability and accuracy in reproducing the complex dynamics of interacting fluid interfaces \cite{gunstensen1991lattice,liu2012three,leclaire2017generalized}.
The aim of the color-gradient model is to modify the collision step to induce the build-up of a surface force at a fluid-fluid interface.
The CG collision can be conveniently splitted into two main steps: perturbation and anti-diffusion.
In the present implementation, the full collision step (relaxation+perturbation) is performed on the total distribution $g=f_R+f_B$ (R and B identifying the interacting red and blue fluids), corresponding to the total density of the fluid ($\rho=\rho_R+\rho_B$) while the anti-diffusion operator operates on the two sets of distributions separately. 
The perturbation step adds the following term to the standard LB collision:

\begin{equation}
    \Omega^1_a= \frac{9}{4} \sigma \omega | \nabla \phi| (t_a \frac{\mathbf{c}_a \cdot \nabla \phi}{\nabla \phi^2} - B_a),
\end{equation}

where $\sigma$ is the surface tension coefficient, $\phi=\frac{\rho_R-\rho_B}{\rho_R+\rho_B}$ is the phase field, whose value ranges between $-1$(pure blue component) and $1$ (pure red), $t_a$ is the set of lattice-dependent weights and $B_a$ is a set of weights needed to recover the correct capillary stress tensor in the momentum equations \cite{liu2012three}. The (non-local) gradient operator appearing in the perturbation term can be efficiently computed by employing discrete gradient formulas on the lattice, as reported in \cite{thampi2013isotropic}.

Once the full relaxation plus perturbation step is performed, an anti-diffusion term is needed to limit the mixing between the interacting immiscible fluids. To this purpose we employ a recolouring term as proposed by Latva and Kokko \cite{latva2005diffusion}:

\begin{equation}
    \Omega^{2}_{a,R}=\frac{\rho_R}{\rho} g_a + \beta \frac{\rho_R\rho_B}{\rho}f_a^{eq,R}(\mathbf{u}=0)cos(\theta),
\end{equation}
\begin{equation}
    \Omega^{2}_{a,B}=\frac{\rho_B}{\rho} g_a + \beta \frac{\rho_R\rho_B}{\rho}f_a^{eq,B}(\mathbf{u}=0)cos(\theta), 
\end{equation}

where $\theta$ is the angle formed between the normal at the interface and the a-th lattice vector, $f^{eq,(R,B)}(\mathbf{u}=0)$ are the (red and blue) equilibrium distribution functions with zero velocity and $\beta$ is a parameter controlling the thickness of the interface. To note, the recoloring is the only operator which needs to be applied on the two sets of distribution separately. 

\subsection{Near-contact interactions for short-range repulsion among fluid interfaces}

Recently, the color-gradient model has been extended to take into account the presence of near-contact repulsive forces between short-range interacting fluid interfaces \cite{montessori2019jfm, montessori2019transa}. The near contact interactions (NCI hereafter) are specifically introduced to frustrate the coalescence between interacting interfaces as occurs in many-body soft systems such as dense emulsions and foams, high internal phase emulsions (HIPEs) and have shown to accurately capture strikingly complex behaviors arising in dense soft interacting systems \cite{bogdan2022stochastic,montessori2021translocation,montessori2019prfluids,montessori2021wet}.

In this work we propose a simpler and more efficient implementation of the NCIs  which avoids explicit ray-tracing along the normal to the fluid interface, thus allowing to keep data locality.
The main ingredients of the new implementation can be sketched as follows: before the collision step, a loop over the fluid nodes is performed to find the interfacial nodes on which the repulsive force will be applied in the subsequent collision step.
This can be done by sitting on a bulk node (i.e., $\phi_{bulk}(i,j,k)< (-1+\epsilon)$, being $\epsilon$ a small number) and checking the presence of two neighboring interfaces along pairs of opposite lattice directions: their presence  can be inferred by sampling the values of $\phi$ along such directions. Thus, if $\phi$, evaluated on points along  opposite lattice directions, is larger than $\phi_{bulk}$, the two points are flagged as eligible to be augmented with NCIs.

In the collision step, an additional force acting on the flagged interfacial points and directed towards the local normal to the interface is added, being such a force proportional to the local value of the dispersed phase density. 
This latter choice permits to take automatically into account the mutual distance among interacting interfaces. Indeed, as two interfaces approach each other, the values of the densities of the dispersed phase, sampled at flagged points, will grow accordingly, as will the repulsive interfacial force.

\section{Thread-safe reconstruction of post-collision distributions } \label{nrpc}

In this section we present the main ideas underlying the thread-safe LB model. The starting point for its derivation is to observe that the post-collision distribution can be rewritten as \cite{succi},
\begin{equation}\label{LBexp}
    f^{pc}_{a} = f_{a} + \omega(f^{eq}_{a} - f_{a})=f^{eq}_{a} + (1-\omega)f^{neq}_{a},
\end{equation}

where the relation $f_a=f^{eq} + f^{neq}$ has been employed.
In other words, $f^{pc}_a$, can be expressed as the sum of its equilibrium value and a non-equilibrium part relaxed at a rate $(1-\omega)$.
Interesting to note that, from an algorithmic standpoint, such an explicit reconstruction decouples (at least theoretically) pre and post-collision values of $f_a$, with evident advantages for LB implementations on shared memory architectures. One for all is that, the above representation of the post-collisional eliminates dependencies in terms of data movement, which are present in fused streaming-collision approaches \cite{mattila, latt2021cross}.
Indeed, the streaming operations can be explicitly included in eq. \ref{LBexp}, as:
\begin{equation}\label{pushLB}
    f_{a}(\mathbf{x}+\mathbf{c}_a,\tau+1) = f^{eq}_{a}(\mathbf{x},t) + (1-\omega)f^{neq}_{a}(\mathbf{x},t).
\end{equation}
The relation reported in eq.\ref{pushLB} becomes a thread-safe operation in shared-memory architecture, as in GPUs, only if the non-equilibrium part can be directly evaluated  by reverse-engineering of the macroscopic quantities, namely density, momentum and momentum flux tensor. Such an operation can be efficiently performed by resorting to the regularization procedure \cite{latt2006}, whose  main 
goal is to convert $f^{neq}_a$ into a new set of non-equilibrium distributions ($\hat{f}^{neq}$) entirely lying on a Hermite subspace spanned by the first three statistical moments only, namely $\rho$, $\rho \mathbf{u}$ and $\Pi^{neq}$ \cite{zhang2006efficient}.
More specifically, by introducing Hermite polynomials and Gauss-Hermite quadratures, $f^{neq}$ can be written as \cite{shan2006kinetic, montessori2015lattice}:

\begin{equation} \label{hermite}
    \hat{f}^{neq}_a=t_a \sum_n \frac{1}{n!}\mathbf{a}^n \mathcal{H}^n(\mathbf{c}_a),
\end{equation}

where $\mathcal{H}^n(\mathbf{c}_a)$ is the standard n-th order Hermite polynomial and $\mathbf{a}^n=\sum_a \hat{f}^{neq}_a \mathcal{H}^n(\mathbf{c}_a)$ the corresponding Hermite expansion coefficient. Both $\mathbf{a}^n$ and  $\mathcal{H}^n(\mathbf{c}_a)$ are n-th rank tensors.

Thus, the non-equilibrium set of distributions, defined as in eq.\ref{hermite}, contains information from hydrodynamic moments up to the order of the expansion ($n=2$, for the Navier-Stokes level), and filters out contributions stemming from higher order fluxes \cite{zhang2006efficient}
The set of non-equilibrium distributions can be compactly written as \cite{latt2006, montessori2015lattice}:

\begin{equation}
    \hat{f}^{neq}_a=\frac{t_a}{2 c_s^4}(\mathbf{c}_a\mathbf{c}_a - c_s^2\delta_{\alpha\beta}):\Pi^{neq}_{\alpha\beta}.
\end{equation}

With the above relation in mind, the lattice kinetic dynamics exposed in eq.\ref{LBexp} can be  implemented by storing, at each time step and lattice node, three macroscopic quantities namely, one scalar ($\rho$), three vector components ($\rho \mathbf{u}$) and six components of a (symmetric), rank two, tensor $\Pi^{neq}_{\alpha\beta}$. 
For the sake of clarity, the full stream and collision step can be written as follows:

\begin{multline}
    f_{a}(\mathbf{x}+\mathbf{c}_a,\tau+1) = t_a\left(\rho + \rho_0\left(\frac{\mathbf{c}_{a} \cdot \mathbf{u}}{c_s^2} + \frac{(\mathbf{c}_a\mathbf{c}_a - c_s^2\mathbb{I}): \mathbf{u}\mathbf{u}}{2c_s^4}   \right )\right) + \\ (1 - \omega)\frac{t_a}{2 c_s^4}(\mathbf{c}_a\mathbf{c}_a - c_s^2\mathbb{I}):\Pi^{neq}
\end{multline}

In other words, the thread-safe LB, rather than explicitly streaming distributions along the lattice directions, aims at reconstructing the post-streamed/post-collided distribution via a scattered reading of the neighboring macroscopic hydrodynamic fields which are then employed to rebuild the full set of distributions. 

\section{Implementation details}

In this section, we detail the main algorithmic  strategies employed to optimize the model's implementation on GPUs.

It is worth noting that, the current implementation makes use of a structure of array (SoA) ordering \cite{shet2013data}. Such choice is motivated since the SoA is necessary to exploit the coalesced memory access, which allows for optimal usage of the global memory bandwidth of the graphic processing device \cite{obrecht2011new}. 
Moreover, all the cycles running through the lattice links have been explicitly unrolled and optimized to avoid the execution of unnecessary floating point operations. 

Without loosing any generality, we proceed with a description of the single component solver and, in the following, we discuss the additional steps to include the color-gradient model.

An LB time step can be split into three major steps, namely moments reconstruction, streaming and collision, and boundary conditions. These are 
recursively performed throughout the entire simulation.

Below we report a code snippet of the moment reconstruction step:

\begin{equation}
\boxed{
\begin{array}{rcl}

 \textrm{for each node (i,j)} \\
 \\
 \rho_{(i,j)}=\sum_a f_{a,(i,j)}\\
 \\
 \rho_{(i,j)}\mathbf{u}_{(i,j)}=\sum_a f_{a,(i,j)}\mathbf{c}_a \\
 \\
  \Pi^{neq}_{\alpha\beta,(i,j)}=\sum_a f^{neq}_{a,(i,j)}c_{a\alpha}c_{a\beta}
\end{array}
}
\end{equation}

Density, linear momentum and pressure tensor components are computed, stored, and subsequently used in the collision step. 
Since the non-equilibrium components of the momentum flux tensor are obtained as $f-f^{eq}$, the equilibrium distributions need to be retrieved in the moment reconstruction step and used to calculate the components $\Pi^{neq}_{\alpha\beta,(i,j)}$ of the non-equilibrium momentum flux tensor. 
It is worth noting here that, the algorithm presented in section \ref{nrpc} needs respectively three (in 2D) and six (in 3D)  additional arrays to be stored ($\Pi^{neq}_{\alpha\beta,(i,j)}$) which are used to perform the reconstruct $f^{neq}_a$ via Hermite projection. 

Nonetheless, it is worth observing that, the memory saving is still significant in both two and three-dimensions, with respect to typical "flip-flop" LB implementation (i.e. an A-B-A scheme \cite{mattila}). Indeed, while a flip-flop LB requires to allocate, roughly, 21 (2D, D2Q9 lattice) and 42 (3D, D3Q19 lattice) arrays, the non-local reconstruction needs, respectively,  15 and 29 arrays to be stored. If the arrays are stored in single precision, as in our GPU implementations, this in turn translates into a memory saving of $\sim 24 \;bytes$ (in 2D) and $\sim 52\; bytes$ (in 3D) per lattice node. By postulating a three-dimensional simulation of $10^9$ computational nodes, this reflects in a net memory saving of $\sim50 \;GBytes$ (roughly the RAM allocated on two Nvidia V100 GPUs or one Nvidia A100). 

As recalled in the introduction, another fundamental feature of the  thread-safe approach lies in its simplicity with respect to other existing single-distribution, in-place streaming approaches, such as EsoTwist \cite{geier2017esoteric, lehmann2022esoteric} and the Bailey algorithm \cite{bailey2009accelerating}, which require the resort to high-level programming language due to the need of explicitly swapping pointers to preserve data locality.
From this standpoint, the thread-safe strategy may be, perspectively, incorporated into existing HPC LB-based software packages.

After the moment reconstruction,  the fused streaming-collision step is performed along the lines drafted in sec.\ref{nrpc}. Below we report a snippet highlighting the main operations involved:

\begin{equation}
\boxed{
\begin{array}{rcl}

 \textrm{for each node (i,j)} \\
uu= \frac{1}{2 c_s^2}(\mathbf{u}_{i,j}\cdot \mathbf{u}_{i,j}) \\
u \cdot c=\frac{\mathbf{c}_{a} \cdot \mathbf{u}_{i,j}}{c_s^2} \\
feq=  t_a \left( \rho_{i,j} + \rho_0\left(u \cdot c + \frac{1}{2}(u \cdot c)^2 - uu \right)\right) \\
fneq=\frac{t_a}{2c_s^2}(c_{a,\alpha}c_{a,\beta} - c_s^2\delta_\alpha\beta):\Pi^{neq}_{\alpha\beta,(i,j)} \\
f_{a,(i+c_{a,x},j+c_{a,y})}=feq + (1-\omega)fneq

\end{array}
}
\end{equation}
 
As recalled in the previous section, the post-collision distributions loose any connection with their pre-collision values. From this standpoint, the streaming step is replaced by a non-local reconstruction of the post-collision set of distributions, performed at neighboring points along the a-th  lattice direction upstream, as evidenced in the snippet above.
This approach is particularly beneficial from a computational standpoint, as it avoids the onset of race conditions that may jeopardize the correct memory access in shared memory GPU architectures. Moreover, the thread-safe reconstruction allows for a straightforward porting of LB codes on GPU, either by resorting to CUDA native language or by employing directive-based procedures, in the second case  with virtually no need to modify the structure of the serial code, as shown below.

In the case of the CG LB, two additional operations have two be taken into account: i) the perturbation step, which is performed on the total distribution function, and ii) the recolouring step, which instead is performed separately on the two sets of distributions. Nonetheless, it is worth recalling that such steps are performed only on a limited subset of lattice nodes belonging to the fluid interface separating the two interacting fluids. Thus, the computational intensity of the perturbation and recolouring steps scales, in three-dimensions, as $O(N^2)$, being $N^3$ the volume of the domain enclosing the multicomponent system.

\subsection{OpenACC implementation}

A first option to port the LB code on GPU is represented by OpenACC (Open Accelerators). OpenACC is a programming standard for parallel computing designed to aid parallel programming on heterogeneous architecture (CPU/GPU).
With OpenACC, the first step is to identify code areas that are prone to be accelerated by annotating the script with directives telling the compiler to perform intensive calculations on the GPU with no need of explicitly manage data transfers between the host and the accelerator.

Below we report a pseudo-code to highlight the calls to OpenACC directives needed to perform the porting of the serial code on GPU.

\begin{lstlisting}

     initialization on host
     
     !***************copy of arrays on device******!
     !$acc data copy(allocated arrays)

     do step=t_in,t_end
     
        !$acc kernels 

        !****moment + neq pressor******************!
         !$acc loop collapse(3)
        do i,j,k
            moments calculations
        enddo

        !****collision +streaming******************!
        !$acc loop collapse(3)
        do i,j,k
            collision+streaming
        enddo
        
        !****boundary conditions******************!
        !$acc loop collapse(2)
        do i,j
            boundary conditions
        enddo
     enddo
    !$acc end data
     
\end{lstlisting}

The advantage of performing a GPU porting via a directive-based approach lies in its simplicity, provided that the serial code is organized in such a way to avoid the presence of processes accessing shared data without explicit synchronization. Moreover, the OpenACC implementation allows for a full portability across operating systems, host CPUs, and a wide range of accelerators, including APUs, GPUs, and many-core coprocessors.
The OpenACC directives employed in the current LB implementation are the following:

\begin{itemize}
    \item \textit{!\$acc data copy} : \textit{data} directive that tells the compiler to create code that performs specific data movements. The \textit{copy} clause, copies data to and from the host and accelerator. When entering the data region, the application allocates accelerator memory and then copies data from the host to the GPU. When exiting the data region, the data from the accelerator is copied back to the host. 

    \item \textit{!\$acc kernels } : The loops nest in a \textit{kernels} construct are converted by the compiler into parallel kernels that run efficiently on your GPU.

    \item \textit{!\$acc loop collapse(n)} : the loop construct applies to the tightly nested loops, and describes the type of device parallelism to use to execute the iterations of the loop. The clause \textit{collapse} applies the associated directive to the following n tightly nested loops.
    
\end{itemize}


\subsection{CUDA Fortran implementation}

Along with the OpenACC version, we also developed  CUDA Fortran implementations of the thread-safe LB approach. In order to keep the comparison between the two GPU codes ( i.e., OpenACC and CUDA Fortran) as fair as possible, the GPU porting with CUDA has been performed by starting from the same serial code employed in the directive-based implementation.

The CUDA porting decomposes the global domain according to a 2D/3D block distribution among the CUDA threads. The selection of the block distribution can be tuned to obtain the best
performance depending on the global grid dimension and the compute
capability of the GPU device. 
In the present implementation, each thread is responsible for one grid point in each CUDA block and the threads iterate over all the populations in the implemented kernels.
Finally, to exploit the memory bandwidth we opted for a decomposition that emphasizes the x-axis dimension due to the data continuity in the column-major order of the FORTRAN language.

Taking into consideration the single component code, three main kernels are called in each time step, namely 

\begin{lstlisting}

initialization on host and on device

do step t_in,t_end

    !****moment + neq pressor******************!
    
    moments<<<dimGrid,dimBlock>>>()

    !****collision +streaming******************!
    
    streamcoll<<<dimGrid,dimBlock>>>()

    !****boundary conditions******************!

    bcs<<<dimGrid,dimBlock>>>()
    
enddo
\end{lstlisting}

\textit{dimGrid} and \textit{dimBlock} parameters are related to the thread abstraction model supported by CUDA. The kernel code will be run by a team of threads in parallel, with the work divided up as specified by the chevron parameters, which are chosen by the user to optimize the performance.

\section{Validation}

The OpenACC and CUDA versions of the thread-safe LB have been fully validated against a  series of benchmarks (both single and multicomponents), namely i) the lid-driven cavity flow, ii) the oscillation of a droplet in a quiescent fluid and iii) the head-on impact between two immiscible droplets. The first test is two-dimensional while the latter two have been performed in three-dimensions. The simulations have been run in single precision on a GPU Nvidia RTX3090.

\subsection{Lid-driven cavity}

\begin{figure}
    \centering
    \includegraphics[scale=0.85]{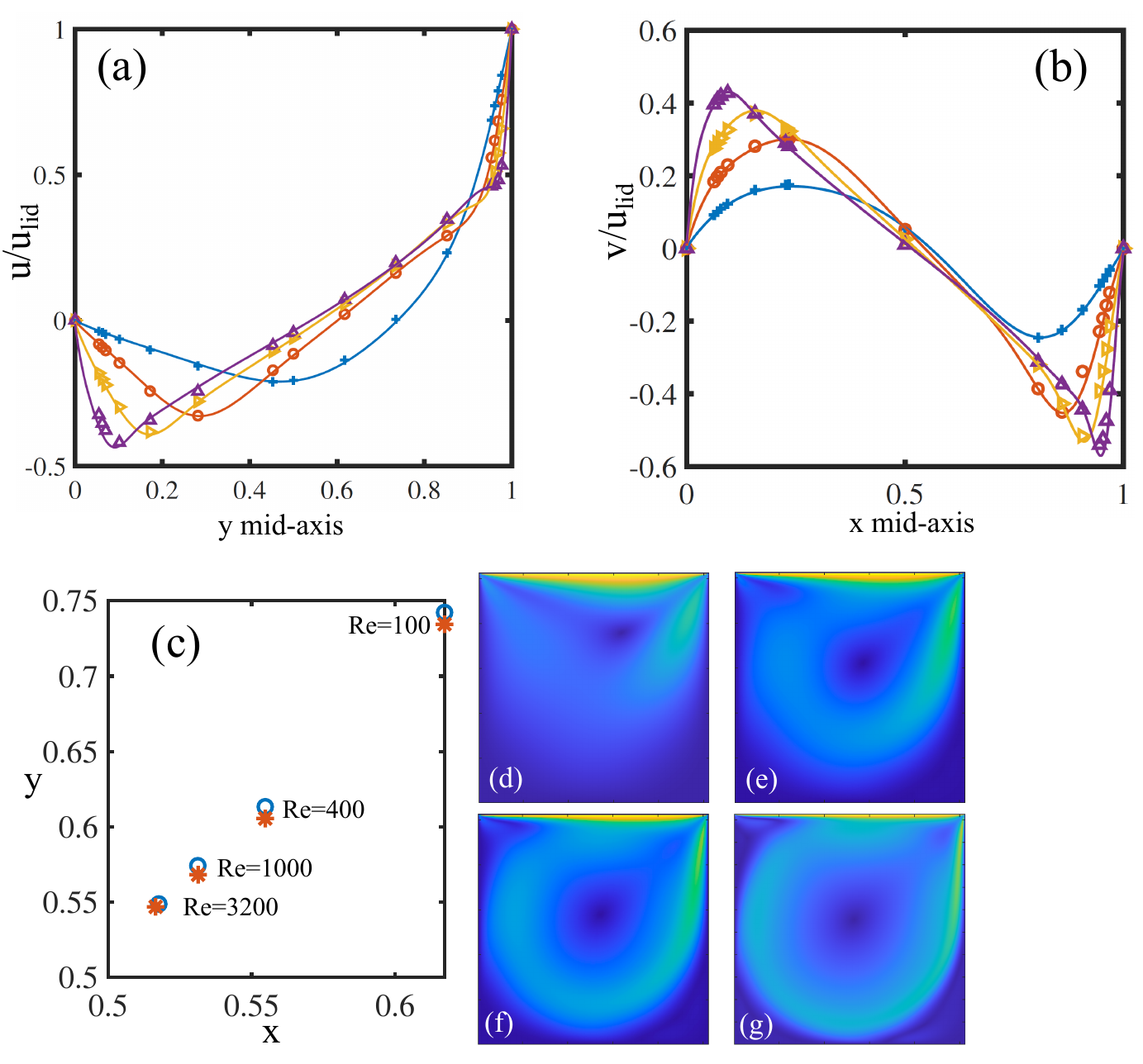}
    \caption{(a) Horizontal velocity profile along the vertical mid-axis of the cavity. (b) Vertical velocity profile along the horizontal mid-axis of the cavity. Symbols refer to reference results \cite{ghia1982high}, solid lines thread-safe LB results. (c) Positions of the vortex center as a function of the Reynolds number. Open circles thread-safe LB, asterisks reference results from \cite{ghia1982high}. (d-g) Steady-state velocity magnitude fields. To note,  the higher accuracy in the prediction of the vortex center at $Re=3200$ is due to the increased resolution employed for this test case.}
    \label{lid driven}
\end{figure}

The lid-driven cavity is the first benchmark we run to test the accuracy of the single-component solver. The main results are reported in figure \ref{lid driven}, panels (a) to (g). Before commenting the results, the main simulation parameters are as follows: the fluid is enclosed in a square box of side $Lx=Ly=256$ ($Re=100-1000$) and $Lx=Ly=512$ ($Re=3200$) and moves due to the motion of the upper lid. The viscosity and lid velocity have been set to $\nu=0.7\;u_{lid}=0.025$ ($Re=100$), $\nu=0.55\;u_{lid}=0.025$ ($Re=400$), $\nu=0.55\;u_{lid}=0.065$ ($Re=1000$), $\nu=0.55\;u_{lid}=0.105$ ($Re=3200$). Bounce-back boundary conditions are imposed at the four walls to enforce no-slip conditions and the correction proposed by Bouzidi hase been implemented to code for the motion of the lid \cite{bouzidi2001momentum}.
Panels (a) and (b) of figure \ref{lid driven} report the horizontal and vertical velocity profiles along the y and x mid-axis of the cavity, respectively. Symbols represent reference data (Ghia et al. \cite{ghia1982high}), while solid lines stand for LB solutions. As one can see, the model reproduces the reference flow profiles with excellent accuracy, within a few percent, for a wide range of Reynolds numbers. The position of the main vortexes at different Reynolds has been inspected as well and compared against those reported in \cite{ghia1982high}. The results, reported in panel (c), confirm the capability of the model to predict the vortex positions for each Reynolds number investigated.

\subsection{Droplet Oscillation in a quiescent fluid}

The multicomponent LB has been first tested against the oscillatory-droplet test. The simulation setup consists of an oblate droplet immersed in a cubic box (whose side is $360$ lattice units) of bulk fluid, which is left free to oscillate and relax toward its spherical shape. The equivalent radius of the droplet is $R_e=40$ lattice units, the bulk and droplet's fluids have the same viscosity, and the surface tension is $\sigma=0.03$ (the data are expressed in lattice units). Gravity is neglected throughout the simulations.
\begin{figure}
    \centering
    \includegraphics[scale=0.65]{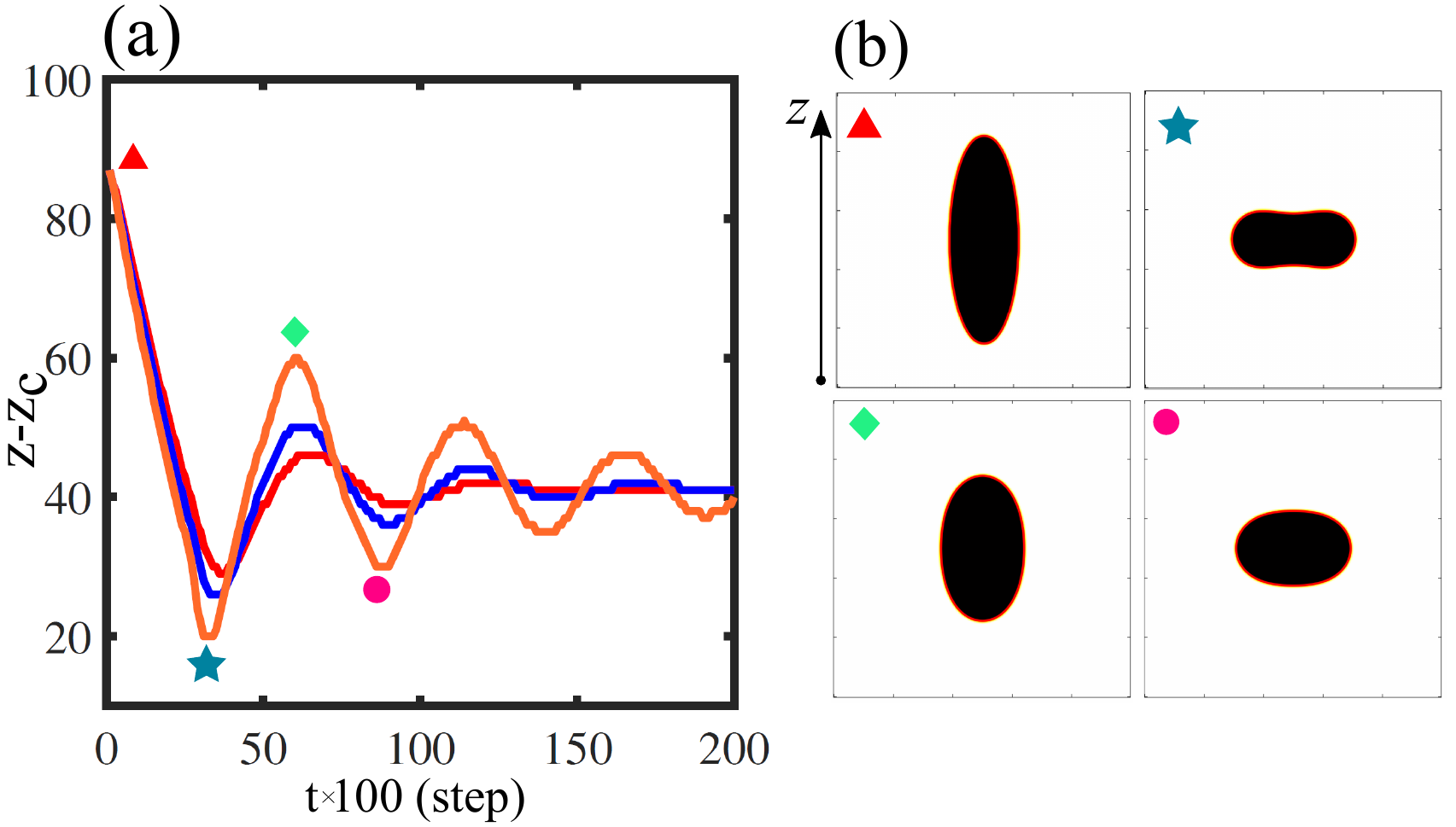}
    \caption{(a) Fluid interface position as a function of time. Simulations have been performed by setting the surface tension $\sigma=0.03$ while the viscosity of the two fluids has been varied as $\nu=0.55$ (orange line), $\nu=0.65$ (blue line), $\nu=0.75$ (red line) (viscosity ratio has been set to unity). The four symbols in panel (a) indicate the four maximum compressions/elongations of the oscillating droplet, as also shown in panel (b) reporting a vertical mid-section of the whole three-dimensional domain .   }
    \label{drop oscillation}
\end{figure}
The solution of Miller and Scriven \cite{miller1968oscillations} for the m-th mode oscillation frequency reads as:

\begin{equation}
    \omega_m=\omega_m^* - \frac{1}{2}\chi\omega_m^{*1/2}+\frac{1}{4}\chi^2
\end{equation}

where $\omega_m$ is the m-th mode of the Lamb frequency \cite{lamb1924hydrodynamics}:
\begin{equation}
\omega_m^*=\sqrt{(m(m+1)(m-1)(m+1))\sigma/(R_e^3(2m+1))}
\end{equation}
. The parameter $\chi$ is defined as:

\begin{equation} \label{chi}
    \chi=\frac{(2n+1)^2(\mu_1\mu_2\rho_1\rho_2)^{1/2}}{2 R_e[n\rho_2+(n+1)\rho_1][(\mu_1\rho_1)^{1/2}+(\mu_2\rho_2)^{1/2}]}
\end{equation}

being $n$ the density ratio (in our simulations is set to unity), $ \mu_1$ and $\mu_2$ the dynamic viscosities of the fluids, $\rho_1$, $\rho_2$ their densities and $R_e$ the equivalent radius of the droplet.
In this work, we are interested in studying the second mode and comparing the theoretical oscillation period ($T_{th}=2 \pi/\omega_2$) with the one obtained via numerical simulations. To do so, we measured the position of the interface along the vertical mid-center line of the cubic box in which the droplet is immersed. One and half relaxation cycle of the droplet is graphically reported in panel (b) of figure \ref{drop oscillation} while, in  panel (a), we reported the time history of the interfacial location for different values of viscosity values.  The numerical oscillation period, computed by considering two subsequent peaks of the signal, agrees well with the theoretical solution (eq. \ref{chi}) with an error $E=|T_{th}-T_{LB}|/T_{th}\sim 2\%$ ($\sigma=0.03$, $\nu=0.75$, $T_{LB}\sim 5400$, $T_{th}=5498$, dimensions expressed in lattice units).

\subsection{Head-on impact between droplets}

In this subsection, we report a qualitative comparison between experiments and LB simulations of head-on impact between two immiscible droplets. The simulation setup consists of two immiscible droplets immersed in a bulk fluid (the two fluids having the same viscosity), which undergo a head-on impact.
\begin{figure}
    \centering
    \includegraphics[scale=1.8]{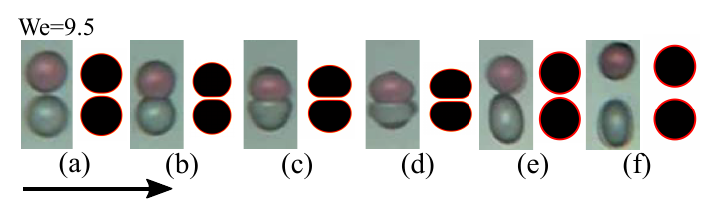}
    \caption{Sequence of the impact between two equally-sized immiscible droplets visualized on a two dimensional mid-plane passing through the droplets centers. Time evolves from left to right as conveniently shown by the black arrow.}
    \label{headon}
\end{figure}

\begin{figure}
    \centering
    \includegraphics[scale=.7]{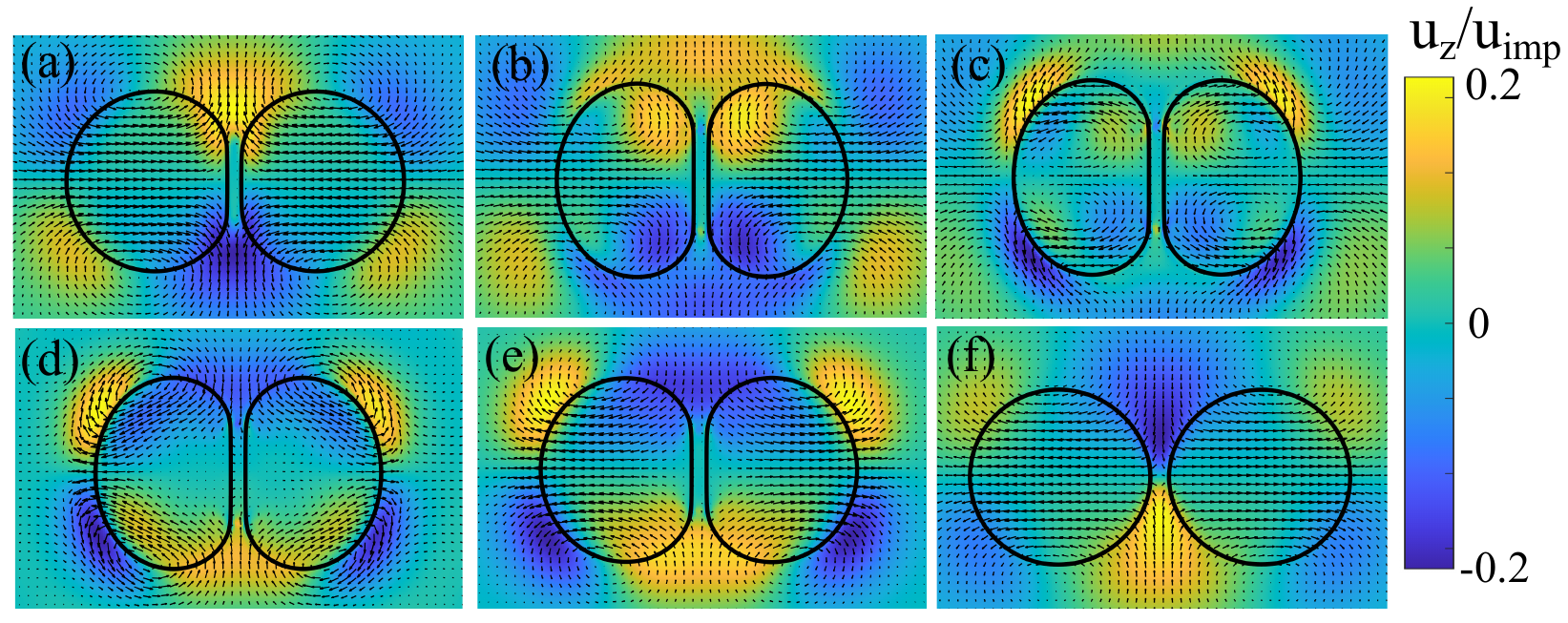}
    \caption{Velocity field during droplet impact. The color maps in panels (a-f) stand for the vertical component of the flow field (values are normalized with the velocity of the droplet at impact) while the superimposed  quivers denote the direction, verse, and magnitude of the local velocity field.  }
    \label{headonvel}
\end{figure}

The main non-dimensional number governing the phenomenon in play is the Weber number ($We=\rho U^2 L/\sigma$) chosen as $We=9.5$, according to the experiments reported in \cite{chen2006collision}. Figure \ref{headon} compares experimental and numerical snapshots taken at different times during the droplet impact. The CGLB equipped with the NCIs is capable of reproducing the overall physical behaviour of the impact, which consists of  i) a  first stage during which the two droplets approach each other and ii) a second stage where the two droplets deform under the effect of the inertia, the maximum deformation depending upon the competition between inertia, near-contact forces (which oppose to interfacial coalescence) and surface tension. The last stage sets the final departure between the two droplets, which separate.
Interesting to note that the lattice Boltzmann simulations give us the possibility to capture important fluid dynamics details occurring during the impact stages. To this aim, figure \ref{headonvel} reports the vertical component of the velocity field (colormap) with the quiver plot superimposed. During the first and second stages (approach and deformation), the fluid is pushed outwards the region comprised within the two droplets. When the droplets reach the maximum deformation, the outward flow first stops (panel (c)) and starts again in the opposite direction when the droplets begin to separate. Such hydrodynamic behaviour is also confirmed by theoretical predictions and observed experimentally, as reported in \cite{kamp2017drop} and references therein.

\section{Performance analysis}

Detailed performance analyses have been performed to test both CUDA and OpenACC implementations of the thread-safe LB model.
The codes have been  run on two different accelerators: Nvidia V100 and GeForce RTX 3090. The former features 5120 Cuda cores and 16 Gbytes RAM, providing a peak performance of 15 TeraFLOPS in single precision with a memory bandwidth of 900 GB/s while, the latter allocates 24 Gbytes of RAM shared by 10496 Cuda cores with a peak performance $\sim 35$ TeraFLOPS in single precision with a memory bandwidth of $\sim 900$ GB/s. 

The performances of the GPU portings have been measured in billions of lattice update per second (GLUPS), defined as:

\begin{equation}
    GLUPS=\frac{n_x n_y n_z n_{steps}}{10^9 \Delta t}
\end{equation}

where $n_{x,y,z}$ are the number of lattice nodes along the three spatial dimensions, $n_{steps}$ is the number of simulation time steps, and $\Delta t$ the run wall-clock time (in seconds) needed to perform the simulation.

The computational performances have been analyzed by choosing square and cubic boxes (in 2d and 3d, respectively) of increasing size, namely $256^2$, $512^2$, $1024^2$, $2048^2$ and $4096^2$ in 2d and $64^3$, $128^3$, $256^3$ and $512^3$ on the RTX 3090 and $416^3$ on the V100 for the single component LB and $64^3$, $128^3$ and $256^3$ for the two-component system. 

\subsection{Single component model}

We start by inspecting the performances of the CUDA and OpenACC versions of the single component LB with thread-safe implementation.\\

\textit{OpenACC implementation}

\begin{figure}
    \centering
    \includegraphics[scale=0.9]{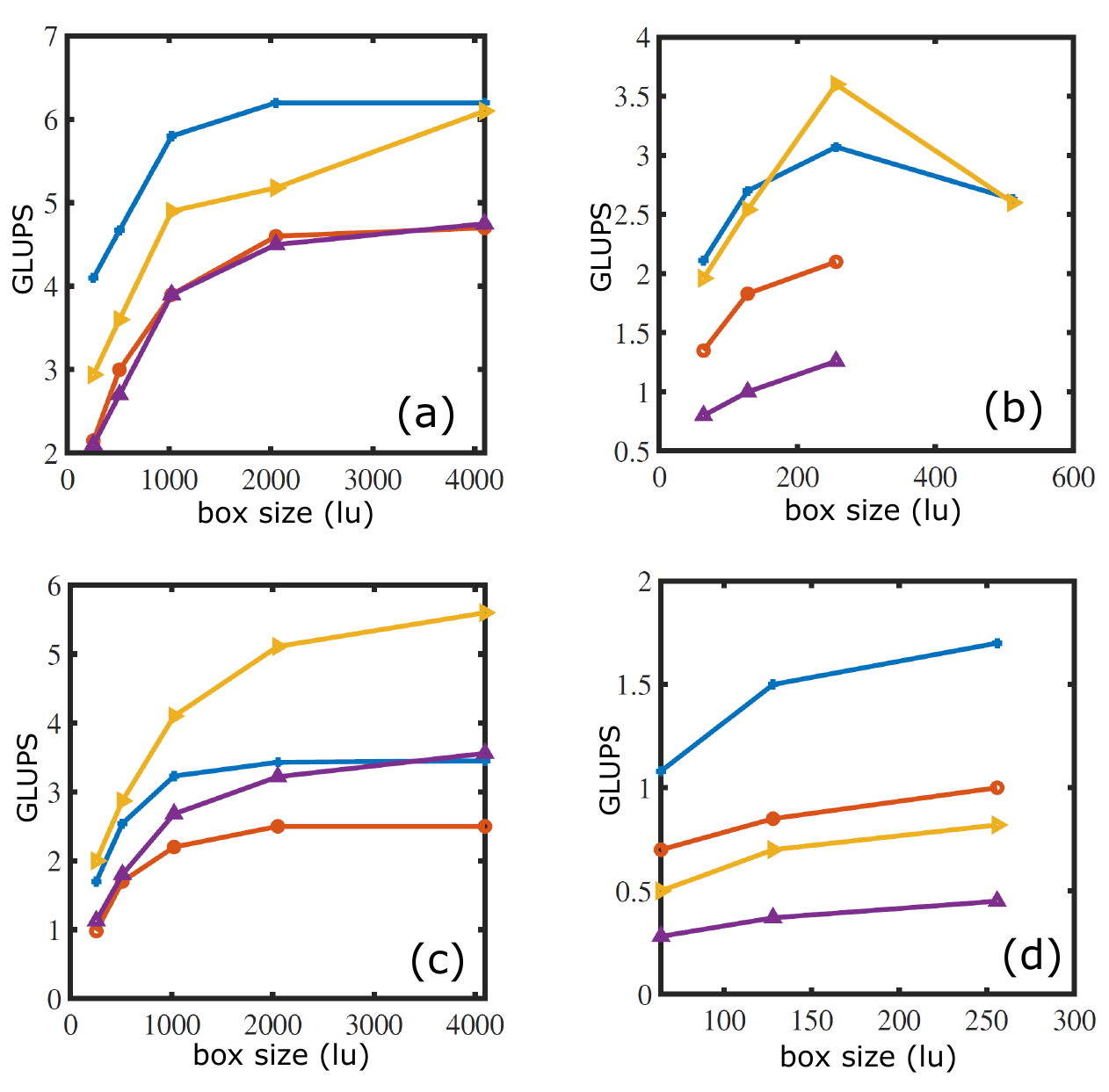}
    \caption{Single and multi-component performance assessment in billions of lattice update per second (GLUPS) as a function of the box size. (a) Single component 2d; (b) single component 3d; (c) two-components 2d; (d) two-components 3d. Blue Line with plusses OpenACC on rtx3090, yellow Line with triangles CUDA on rtx3090, red line with circles OpenACC on V100, violet line with triangles CUDA on v100. }
    \label{fig:perf}
\end{figure}

Let us start with the directive-based version of the code. As one can see from figure \ref{fig:perf} the peak performance in 2D is reached for a square box of side $2048^2$, and the code delivers  $4.6$ and $6.2$ GLUPS on V100 and RTX 3090, respectively. Increasing the size to $nx=ny=nz=4096$ does not lead to any significant improvement in the performances. In 3D, the optimal case is represented by a cubic box of side $256^3$ where the code delivers $2.1$ and $3.1$ GLUPS (V100 and RTX3090). 


We highlight that, in absolute terms, such performances are in line with other HPC (CUDA native) implementations such as those reported in \cite{bonaccorso2022lbcuda, lehmann2022esoteric, latt2021palabos}. However, it is worth noting that, while the cited implementations are based on standard BGK collisional operator, our approach exploits the explicit projections onto (full hydrodynamic) Hermite subspaces, a procedure that filters out the (non-physical) ghost modes from pre-collision distributions. From this standpoint, the LB used in this work is a more accurate and stable (thus allowing to span a broader range of viscosities \cite{montessori2014regularized}) but even more computationally intensive alternative to the standard BGK model \cite{latt2006, montessori2014regularized}, since it requires a larger number of FPOs  to be performed per lattice update. Nevertheless, the thread-safe LB delivers performances in line with state-of-the-art solvers with much-reduced occupancy in terms of memory allocation. In passing, it is interesting to note that such peak performances have been obtained without any extra cost in terms of coding due to the possibility to handle GPU computations via OpenACC annotations.\\

\textit{CUDA Fortran implementation}

The CUDA porting of the single component code with thread-safe reconstruction delivers peak performances of $6.1$ (RTX3090) and $4.75$ (V100) GLUPS in the 2d case for a square box of side $nx=4096$ while, in 3D (box size $256^3$), the CUDA porting provides $3.6$ (RTX3090) and $1.3$ GLUPS (V100).

Thus, the two implementations deliver approximately the same performances as also testified by the roofline models reported in subsection \cite{exasc2}.
From the above considerations, the openACC alternative seems to be a viable solution to obtain scalable and portable codes with virtually no extra cost in terms of coding complexity. 
As a concluding remark, we point out that the LB implementations (both directive-based and CUDA) with fused streaming and standard BGK collision with flip-flop strategy (two distributions) deliver a peak performance of 9 GLUPS in 2D and $\sim 5$ in 3D on RTX 3090. Such performances, in line with state-of-the-art LB codes, come with an extra cost in terms of allocated memory on the GPU since the flip-flop strategy requires the allocation of two sets of distributions to avoid race conditions in the streaming step. 
As we shall see by inspecting the roofline model, the thread-safe LB model is capable of delivering peak performances, limited by the relatively low operational intensity of the LB model, in both two and three dimensions.  On the other hand it is interesting to note that the use of a hybrid floating point precision may be the key to exploit in full the memory bandwidth of the GPUs and future investigations will be directed along this lines.

\subsection{Two components model}

\textit{OpenACC implementation}

The openACC implementation of the CGLB model for two-component flows with NCIs delivers the following peak performances (see figure \ref{fig:perf}):

- 2D, square box size $4096^3$ $\sim 3.4$ GLUPS on rtx3090 and $\sim 2.5$ on V100.

- 3D, cubic box size $256^3$ $\sim 1.7$ GLUPS on rtx 3090 and $\sim 1$ on V100.

In this case, the performances are lowered due to the increased computational intensity of the CGLB. Indeed, at variance with the single component model, the CGLB needs i) to evolve two sets of distributions and ii) to perform a larger number of FPOs to update a lattice node. To note, we reduce the FPOs per lattice update by exploiting the following strategies: i) the relaxation+perturbation step is performed on the total distribution function while ii) the recoloring step and the fused streaming push are operated on the two sets of distributions separately. 
Even in this case, the present implementation behaves better in terms of computational speed, being faster (of a factor $\sim 2-2.5$) compared to previous state-of-the-art implementations of CGLB based on single relaxation time collision operator on single GPU \cite{bonaccorso2022lbcuda}.

\textit{CUDA Fortran implementation}

As in the single component case, the peak performance of the CUDA Fortran implementation of the thread-safe CG LB with NCIs are comparable to the directive-based version of the code:

- 2D, square box size $4096^2$ $\sim 5.6$ GLUPS on RTX3090 and $\sim 3.56$ GLUPs on V100.

- 3D, cubic box size $256^3$ $\sim 0.8$ GLUPS on RTX3090 and $\sim 0.5$ GLUPs on V100.

The above results suggest that the explicit execution of nested for loops in the 3D cuda code is the main cause of performance loss. Indeed, with OpenACC, it is possible to exploit the \textit{collapse} directive to merge two or more large loops, to achieve a higher level of parallelism and to optimize data access. More investigation on such fine-level optimizations is needed and will be the subject of future work. A Detailed report of the performance tests described so far is conveniently shown in figure \ref{fig:perf}. 

\subsection{Rooftop models and computational intensities}

In this subsection, we inspect the performances of the OpenACC and CUDA Fortran implementations of the thread-safe LB by inspecting the roofline model. A roofline model is an intuitive approach that can be used to infer the performances of a compute kernel running on multi-core, many-core, and accelerator-based architectures.
\begin{figure}
\centering
    \includegraphics[scale=1.4]{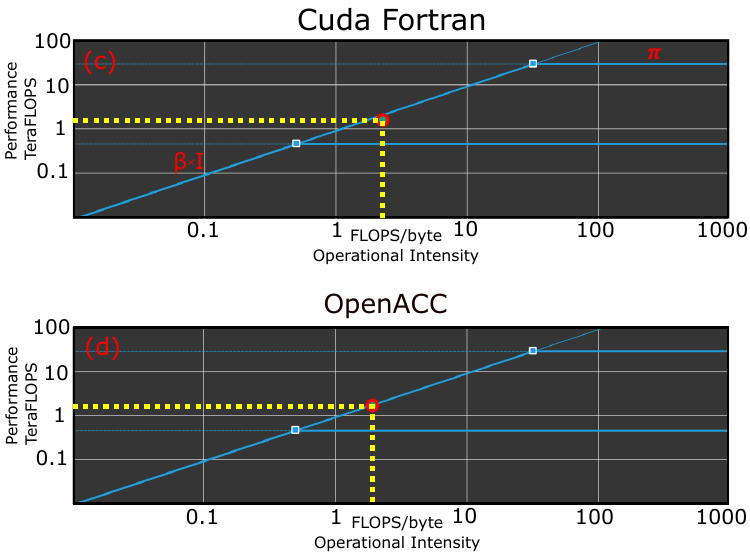}
    \caption{Rooftop models for CUDA and OpenACC implementations of single component thread-safe LB stream and collision subroutine.  The thread-safe LB delivers peak performance of roughly $1.5 TeraFLOP/s$ corresponding to an operational intensity of $\sim 2.31 FLOP/byte$ (CUDA version) and $1.6 TeraFLOPS$ versus an operational intensity of $\sim 2 FLOPS/byte$ }
    \label{fig:roof}
\end{figure}
The roofline model can be visualized by plotting the floating point performance (the work expressed in Floating Point Operation per Second, i.e., FLOP/s or FLOPS) as a function of the operational intensity, as measured in FLOPS/byte, namely the ratio between the work and the memory traffic which denotes the number of bytes of memory transfers occurring during the execution of a kernel. The resultant limit curve is an effective performance bound which sets the performance limit of the kernel under scrutiny. Such a curve includes two platform-specific performance ceilings (the roofs), one which depends on the memory bandwidth while the other is derived from the processor's peak performance.
The attainable performance, namely the roofs in the roofline model, can be derived by the following formula:
\begin{equation}
    P=min(\pi,\beta\times I)
\end{equation}

where $\pi$ is the peak performance, $\beta$ is the peak bandwidth and $I$ the operational intensity (see fig. \ref{fig:roof}). 
The ridge point, namely the point where the two ceilings meet, provides the minimum arithmetic intensity required to be able to achieve the peak performance $\pi$.

As a last remark, since a kernel is characterized by its own arithmetic intensity, its attainable performance $P$ can be assessed by drawing a vertical line hitting the roof. The kernel is said to be memory-bounded if $I<=\pi/\beta$ else the kernel is said to be compute-bound.

In figure \ref{fig:roof} roofline plots for both CUDA and OpenACC implementations of three-dimensional, single component, thread-safe streaming/collision kernels are reported. The kernels  expose an operational intensity of roughly $2.31$ (CUDA) and $2$ (OpenACC) FLOPS/byte, in line with those observed in other state-of-the-art LB implementations based on BGK collision,  achieving almost ideal peak performance of $\sim 1.5 \; TeraFlop/s$ (CUDA) and $\sim 1.6 \; TeraFlop/s$ (OpenACC)  on single GPU (RTX 3090), as evidenced by the hollow red circles sitting on the memory bandwidth boundary $\beta \times I$. The OpenACC implementation delivers a slightly better performance if compared to the native CUDA implementation, as can be  inferred from the plots in fig. \ref{fig:roof}. It is worth noting that, the use of mixed precision \cite{lehmann2022accuracy} would allow to increase the operational intensity of the present codes thus permitting to shift the performances along the $\beta\times I$, for example by using half precision to store fluid populations while still carrying arithmetic operations in single precision, so to optimize the usage of the GPU memory bandwidth.

\section{Conclusions}

In this work, we presented a thread-safe, HPC implementation of the lattice Boltzmann method specifically aimed at exploiting the memory bandwidth of GPU-based architectures. The OpenACC and  CUDA Fortran codes have been shown to provide outstanding performances with much-reduced memory occupancy with respect to fused streaming-collision strategies, commonly employed in high-performance computing lattice Boltzmann codes, and without the need to resort to more complex and exotic streaming algorithms such as those reported in \cite{lehmann2022esoteric, bailey2009accelerating,geier2017esoteric}. The intrinsic simplicity of the thread-safe LB makes it a good candidate for its implementation in existing HPC LB-based codes.
Future works will be focused on assessing the scalability of the thread-safe LB on multi-GPU architectures and on assessing the possibility to employ mixed precisions to optimize the use of the GPU memory bandwidths. 

\section*{Acknowledgements}

M.L. acknowledge funding from MIUR under the project “3D-Phys” (Grant No. PRIN 2017PHRM8X). 
S.S., M.D. acknowledge funding from the European Research Council under the European Union's Horizon 2020 Framework Programme (No.
FP/2014-2020) ERC Grant Agreement No.739964 (COPMAT). 

\label{}

\bibliographystyle{elsarticle-num}

\begin{thebibliography}{10}
\expandafter\ifx\csname url\endcsname\relax
  \def\url#1{\texttt{#1}}\fi
\expandafter\ifx\csname urlprefix\endcsname\relax\def\urlprefix{URL }\fi
\expandafter\ifx\csname href\endcsname\relax
  \def\href#1#2{#2} \def\path#1{#1}\fi

\bibitem{succi}
S.~Succi, The Lattice Boltzmann equation: For complex states of flowing matter,
  Oxford University Press, 2018.

\bibitem{exasc2}
S.~Succi, G.~Amati, M.~Bernaschi, G.~Falcucci, M.~Lauricella, A.~Montessori,
  Towards exascale lattice boltzmann computing, Computer \& Fluids 181 (2019)
  107--115.

\bibitem{bailey2009accelerating}
P.~Bailey, J.~Myre, S.~D. Walsh, D.~J. Lilja, M.~O. Saar, Accelerating lattice
  boltzmann fluid flow simulations using graphics processors, in: 2009
  international conference on parallel processing, IEEE, 2009, pp. 550--557.

\bibitem{geier2017esoteric}
M.~Geier, M.~Sch{\"o}nherr, Esoteric twist: An efficient in-place streaming
  algorithmus for the lattice boltzmann method on massively parallel hardware,
  Computation 5~(2) (2017) 19.

\bibitem{lehmann2022esoteric}
M.~Lehmann, Esoteric pull and esoteric push: Two simple in-place streaming
  schemes for the lattice boltzmann method on gpus, Computation 10~(6) (2022)
  92.

\bibitem{zhang2006efficient}
R.~Zhang, X.~Shan, H.~Chen, Efficient kinetic method for fluid simulation
  beyond the navier-stokes equation, Physical Review E 74~(4) (2006) 046703.

\bibitem{succi2018}
S.~Succi, The Lattice Boltzmann equation: For complex states of flowing matter,
  Oxford University Press, 2018.

\bibitem{kruger2017lattice}
T.~Kr{\"u}ger, H.~Kusumaatmaja, A.~Kuzmin, O.~Shardt, G.~Silva, E.~M. Viggen,
  The lattice boltzmann method, Springer International Publishing 10~(978-3)
  (2017) 4--15.

\bibitem{montessori2018lattice}
A.~Montessori, G.~Falcucci, Lattice Boltzmann modeling of complex flows for
  engineering applications, Morgan \& Claypool Publishers, 2018.

\bibitem{montessori2015lattice}
A.~Montessori, P.~Prestininzi, M.~La~Rocca, S.~Succi, Lattice boltzmann
  approach for complex nonequilibrium flows, Physical Review E 92~(4) (2015)
  043308.

\bibitem{he1997lattice}
X.~He, L.-S. Luo, Lattice boltzmann model for the incompressible navier--stokes
  equation, Journal of statistical Physics 88 (1997) 927--944.

\bibitem{chapman1990mathematical}
S.~Chapman, T.~G. Cowling, The mathematical theory of non-uniform gases: an
  account of the kinetic theory of viscosity, thermal conduction and diffusion
  in gases, Cambridge university press, 1990.

\bibitem{shan1993lattice}
X.~Shan, H.~Chen, Lattice boltzmann model for simulating flows with multiple
  phases and components, Physical review E 47~(3) (1993) 1815.

\bibitem{swift1996lattice}
M.~R. Swift, E.~Orlandini, W.~Osborn, J.~Yeomans, Lattice boltzmann simulations
  of liquid-gas and binary fluid systems, Physical Review E 54~(5) (1996) 5041.

\bibitem{montessori2017entropic}
A.~Montessori, P.~Prestininzi, M.~La~Rocca, S.~Succi, Entropic lattice
  pseudo-potentials for multiphase flow simulations at high weber and reynolds
  numbers, Physics of Fluids 29~(9) (2017) 092103.

\bibitem{benzi2006mesoscopic}
R.~Benzi, L.~Biferale, M.~Sbragaglia, S.~Succi, F.~Toschi, Mesoscopic two-phase
  model for describing apparent slip in micro-channel flows, Europhysics
  letters 74~(4) (2006) 651.

\bibitem{anderl2014free}
D.~Anderl, S.~Bogner, C.~Rauh, U.~R{\"u}de, A.~Delgado, Free surface lattice
  boltzmann with enhanced bubble model, Computers \& Mathematics with
  Applications 67~(2) (2014) 331--339.

\bibitem{thommes2009lattice}
G.~Th{\"o}mmes, J.~Becker, M.~Junk, A.~K. Vaikuntam, D.~Kehrwald, A.~Klar,
  K.~Steiner, A.~Wiegmann, A lattice boltzmann method for immiscible multiphase
  flow simulations using the level set method, Journal of Computational Physics
  228~(4) (2009) 1139--1156.

\bibitem{gunstensen1991lattice}
A.~K. Gunstensen, D.~H. Rothman, S.~Zaleski, G.~Zanetti, Lattice boltzmann
  model of immiscible fluids, Physical Review A 43~(8) (1991) 4320.

\bibitem{liu2012three}
H.~Liu, A.~J. Valocchi, Q.~Kang, Three-dimensional lattice boltzmann model for
  immiscible two-phase flow simulations, Physical Review E 85~(4) (2012)
  046309.

\bibitem{leclaire2017generalized}
S.~Leclaire, A.~Parmigiani, O.~Malaspinas, B.~Chopard, J.~Latt, Generalized
  three-dimensional lattice boltzmann color-gradient method for immiscible
  two-phase pore-scale imbibition and drainage in porous media, Physical Review
  E 95~(3) (2017) 033306.

\bibitem{thampi2013isotropic}
S.~P. Thampi, S.~Ansumali, R.~Adhikari, S.~Succi, Isotropic discrete laplacian
  operators from lattice hydrodynamics, Journal of Computational Physics 234
  (2013) 1--7.

\bibitem{latva2005diffusion}
M.~Latva-Kokko, D.~H. Rothman, Diffusion properties of gradient-based lattice
  boltzmann models of immiscible fluids, Physical Review E 71~(5) (2005)
  056702.

\bibitem{montessori2019jfm}
A.~Montessori, M.~Lauricella, N.~Tirelli, S.~Succi, Mesoscale modelling of
  near-contact interactions for complex flowing interfaces, Journal of Fluid
  Mechanics 872 (2019) 327--347.

\bibitem{montessori2019transa}
A.~Montessori, M.~Lauricella, S.~Succi, Mesoscale modelling of soft flowing
  crystals, Philosophical Transactions of the Royal Society A 377~(2142) (2019)
  20180149.

\bibitem{bogdan2022stochastic}
M.~Bogdan, A.~Montessori, A.~Tiribocchi, F.~Bonaccorso, M.~Lauricella,
  L.~Jurkiewicz, S.~Succi, J.~Guzowski, Stochastic jetting and dripping in
  confined soft granular flows, Physical Review Letters 128~(12) (2022) 128001.

\bibitem{montessori2021translocation}
A.~Montessori, A.~Tiribocchi, M.~Bogdan, F.~Bonaccorso, M.~Lauricella,
  J.~Guzowski, S.~Succi, Translocation dynamics of high-internal phase double
  emulsions in narrow channels, Langmuir 37~(30) (2021) 9026--9033.

\bibitem{montessori2019prfluids}
A.~Montessori, M.~Lauricella, A.~Tiribocchi, S.~Succi, Modeling pattern
  formation in soft flowing crystals, Physical Review Fluids 4~(7) (2019)
  072201.

\bibitem{montessori2021wet}
A.~Montessori, A.~Tiribocchi, M.~Lauricella, F.~Bonaccorso, S.~Succi, Wet to
  dry self-transitions in dense emulsions: From order to disorder and back,
  Physical Review Fluids 6~(2) (2021) 023606.

\bibitem{mattila}
K.~Mattila, J.~Hyv\"aluoma, J.~Timonen, T.~Rossi, Comparison of implementations
  of the lattice-boltzmann method, Computers \& Mathematics with Applications
  55 (2008) 1514--1524.

\bibitem{latt2021cross}
J.~Latt, C.~Coreixas, J.~Beny, Cross-platform programming model for many-core
  lattice boltzmann simulations, Plos one 16~(4) (2021) e0250306.

\bibitem{latt2006}
J.~Latt, B.~Chopard, Lattice boltzmann method with regularized pre-collision
  distribution functions, Mathematics and Computers in Simulation 72 (2006)
  165--168.

\bibitem{shan2006kinetic}
X.~Shan, X.-F. Yuan, H.~Chen, Kinetic theory representation of hydrodynamics: a
  way beyond the navier--stokes equation, Journal of Fluid Mechanics 550 (2006)
  413--441.

\bibitem{shet2013data}
A.~G. Shet, S.~H. Sorathiya, S.~Krithivasan, A.~M. Deshpande, B.~Kaul, S.~D.
  Sherlekar, S.~Ansumali, Data structure and movement for lattice-based
  simulations, Physical Review E 88~(1) (2013) 013314.

\bibitem{obrecht2011new}
C.~Obrecht, F.~Kuznik, B.~Tourancheau, J.-J. Roux, A new approach to the
  lattice boltzmann method for graphics processing units, Computers \&
  Mathematics with Applications 61~(12) (2011) 3628--3638.

\bibitem{bouzidi2001momentum}
M.~Bouzidi, M.~Firdaouss, P.~Lallemand, Momentum transfer of a
  boltzmann-lattice fluid with boundaries, Physics of fluids 13~(11) (2001)
  3452--3459.

\bibitem{ghia1982high}
U.~Ghia, K.~N. Ghia, C.~Shin, High-re solutions for incompressible flow using
  the navier-stokes equations and a multigrid method, Journal of computational
  physics 48~(3) (1982) 387--411.

\bibitem{miller1968oscillations}
C.~Miller, L.~Scriven, The oscillations of a fluid droplet immersed in another
  fluid, Journal of fluid mechanics 32~(3) (1968) 417--435.

\bibitem{lamb1924hydrodynamics}
H.~Lamb, Hydrodynamics, University Press, 1924.

\bibitem{chen2006collision}
R.-H. Chen, C.-T. Chen, Collision between immiscible drops with large surface
  tension difference: diesel oil and water, Experiments in fluids 41 (2006)
  453--461.

\bibitem{kamp2017drop}
J.~Kamp, J.~Villwock, M.~Kraume, Drop coalescence in technical liquid/liquid
  applications: A review on experimental techniques and modeling approaches,
  Reviews in Chemical Engineering 33~(1) (2017) 1--47.

\bibitem{bonaccorso2022lbcuda}
F.~Bonaccorso, M.~Lauricella, A.~Montessori, G.~Amati, M.~Bernaschi, F.~Spiga,
  A.~Tiribocchi, S.~Succi, Lbcuda: A high-performance cuda port of lbsoft for
  simulation of colloidal systems, Computer Physics Communications 277 (2022)
  108380.

\bibitem{latt2021palabos}
J.~Latt, O.~Malaspinas, D.~Kontaxakis, A.~Parmigiani, D.~Lagrava, F.~Brogi,
  M.~B. Belgacem, Y.~Thorimbert, S.~Leclaire, S.~Li, et~al., Palabos: parallel
  lattice boltzmann solver, Computers \& Mathematics with Applications 81
  (2021) 334--350.

\bibitem{montessori2014regularized}
A.~Montessori, G.~Falcucci, P.~Prestininzi, M.~La~Rocca, S.~Succi, Regularized
  lattice bhatnagar-gross-krook model for two-and three-dimensional cavity flow
  simulations, Physical Review E 89~(5) (2014) 053317.

\bibitem{lehmann2022accuracy}
M.~Lehmann, M.~J. Krause, G.~Amati, M.~Sega, J.~Harting, S.~Gekle, Accuracy and
  performance of the lattice boltzmann method with 64-bit, 32-bit, and
  customized 16-bit number formats, Physical Review E 106~(1) (2022) 015308.

\end{thebibliography}

\end{document}